\documentclass{article}

\usepackage[english]{babel}
\usepackage[letterpaper, top=2cm, bottom=2cm, left=3cm, right=3cm, marginparwidth=1.75cm]{geometry}

\usepackage{amsmath}
\usepackage{graphicx}
\usepackage{authblk}
\usepackage{xcolor}

\usepackage{cite}

\usepackage{lineno}

\title{Evolution of phenocopying in a dynamical model of developmental trajectories}
\author[1]{Yuuki Matsushita}
\author[1]{Archishman Raju
\thanks{archishman@ncbs.res.in}}
\affil[1]{National Centre for Biological Sciences, Tata Institute of Fundamental Research, Bangalore, India}

\begin{document}
      \maketitle

      \subsubsection*{Abstract}
      Developmental trajectories are known to be canalized, or robust to both environmental and genetic perturbations. However, even when these trajectories are decanalized by an environmental perturbation outside of the range of conditions to which they are robust, they often produce phenotypes similar to known mutants called phenocopies. This correspondence between the effect of environmental and genetic perturbations has received little theoretical attention. Here, we study an abstract regulatory model which is evolved to follow a specific trajectory. We then study the effect of both small and large perturbations to the trajectory both by changing parameters and by perturbing the state in a timed manner. We find, surprisingly, that the phenomenon of phenocopying emerges in evolved trajectories even though the alternative trajectories are not selected for. Our results suggest that evolution simplifies the structure of high-dimensional phenotypic landscapes which can simultaneously show robustness and phenocopying.




      \subsubsection*{Introduction}

      A central property of organismal development is that of canalization, or robustness to both genetic and environmental perturbations. This robustness has been both experimentally and theoretically studied~\cite{scharloo1991canalization, wagner1997population, flatt2005evolutionary}. Early experiments by Waddington, Rendel and others investigated the nature of canalization for different characters in the fly~\cite{Waddington1959-qj, Waddington1961-oh, Rendel1959-wv}. Subsequent studies have found that canalization may be a property of gene regulatory networks~\cite{barkai2007variability, von2000segment} while others have suggested that so-called capacitors like Hsp90 may play an important role~\cite{Rutherford1998-hh,Queitsch2002-tg,Yeyati2007-fj}. The computational studies on canalization have focused more on genetic canalization and showed that robustness can evolve as a property in more abstract regulatory networks~\cite{Wagner1996-lo,Wagner2000-gj, Siegal2002-bq, Masel2004-ky}. Typically, such studies focus on the robustness of the final phenotype to mutations.

      However, developmental trajectories often show specific and reproducible outcomes even on being environmentally perturbed beyond their normal range of robustness. It has long been known that environmental perturbations can phenocopy known genetic mutants. The term phenocopy was coined by Goldschmidt who initiated a rich experimental literature on the topic~\cite{Goldschmidt1949-jj, goldschmidt1956new, rapoport1947synthesis, sang1954production, lambert1989phenocopies}. Several different kinds of environmental perturbations were shown to produce phenocopies including both heat and cold shocks, as well as exposure to chemicals like ether or salts like silver nitrate. Typically, these perturbations were given at a specific stage of development, were well outside the normal ranges of fluctuations that the organism was likely to be exposed to, and significantly decreased organismal viability. In \textit{Drosophila}, they produced changes in bristles, wings, eyes etc. Some of these were minor alterations like a break in the posterior cross-vein while others like the conversion of halteres into wings were more dramatic.

      The phenomenon of phenocopies suggests that a robust developmental trajectory can be pushed to an alternative one by a large enough perturbation. Further, the experiments demonstrated that the alternative trajectory chosen is very sensitive to the timing of the perturbation. Indeed, different phenocopies in the fly can be obtained simply by adjusting the timing of a heat shock, which was carefully studied in subsequent works~\cite{Milkman1966-fp, Mitchell1966-vo, Petersen1990-pf, Petersen1991-no}.

      In contrast to robustness, the phenomenon of phenocopies has received almost no theoretical attention. Fundamentally, it indicates that the action of an external environment on the state of the system at particular times can mimic changes to its underlying constituents or parameters. This observation suggests that the developmental landscape has a complex but very specific structure. It is not known what kinds of properties are required to reproduce this phenomenon.

      Waddington used the term ``homeorhesis" to describe entire dynamical trajectories that were robust to perturbations during development~\cite{Waddington2014-yq}. Several recent studies have therefore emphasized that development should be understood using concepts from dynamical systems theory~\cite{Rand2021-rz,Matsushita2022-ol, Raju2023-zf}. The dynamical structure of development is crucial to understand and interpret phenocopies for the reasons outlined above.

      Here we computationally study an abstract regulatory model evolved to follow a reference trajectory. In particular, we study whether different kinds of perturbations have similar effects on an evolved population and whether canalization and phenocopying can be obtained as emergent properties of this evolution. To demonstrate the effect of evolution, we compare with randomly sampled populations which also follow the reference trajectory and provide a null model for comparison. Random sampling allows us to sample functional parameters without evolutionary biases.

      The paper is organized as follows. We first describe our dynamical systems model and optimize it to achieve a desired trajectory. To avoid direct identification with molecular components, we investigate the effect of ``internal" and ``external" perturbations. The first of these changes the parameters of the model while the latter changes the trajectory in a timed manner. We describe how the property of robustness and phenocopying emerge from our evolutionary simulations and end by offering some preliminary explanations for this from the point of view of dynamical systems theory.

      \subsubsection*{Model}

\begin{figure}[htbp]
            \centering
            \begin{minipage}[c]{0.25\linewidth}
                  \begin{flushleft}
                        A
                  \end{flushleft}
                  \centering
                  \includegraphics[width=0.6\textwidth]{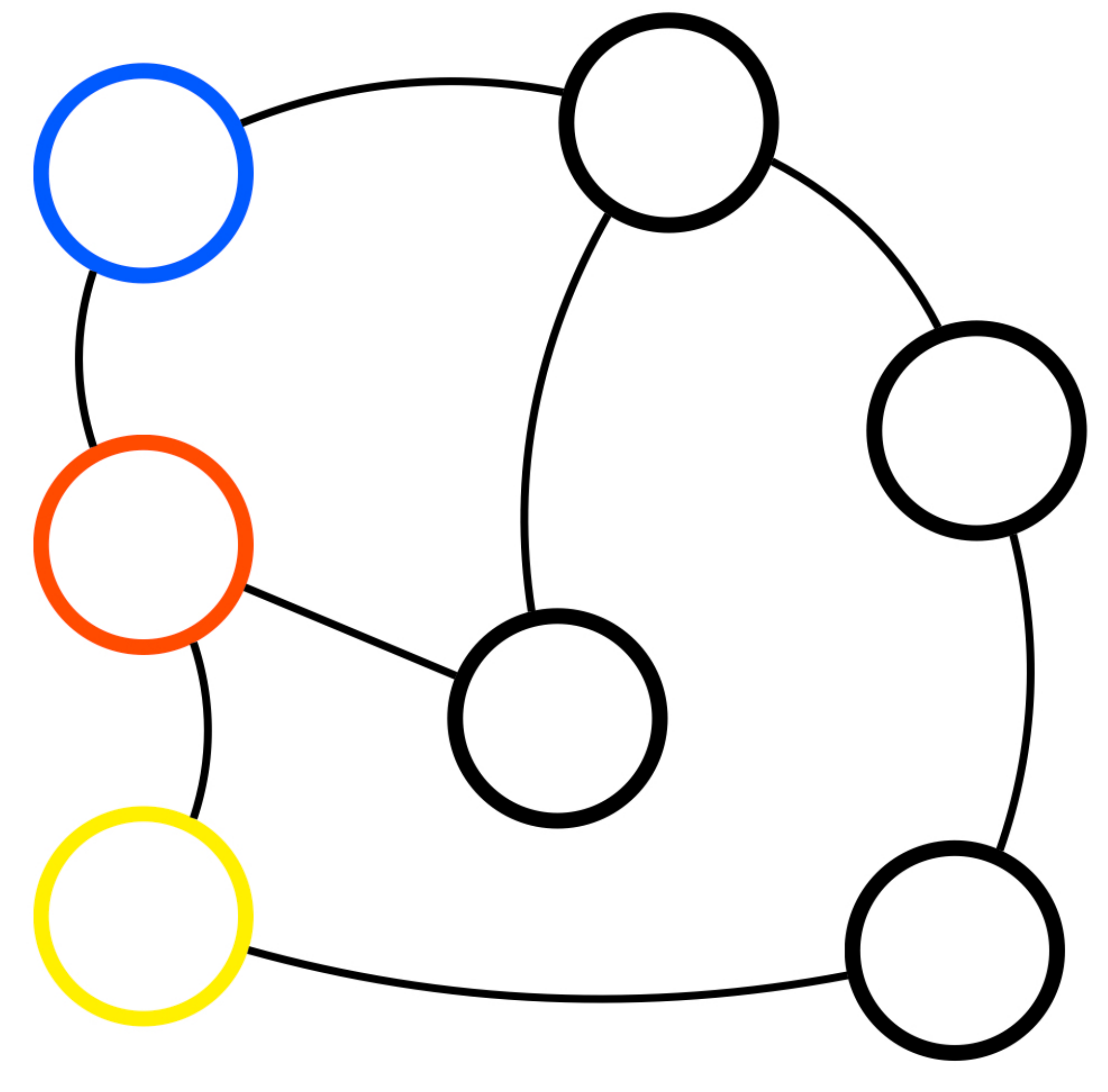}
            \end{minipage}
            \begin{minipage}[c]{0.36\linewidth}
                  \centering
                  \begin{flushleft}
                        B
                  \end{flushleft}
                  \includegraphics[width=\textwidth]{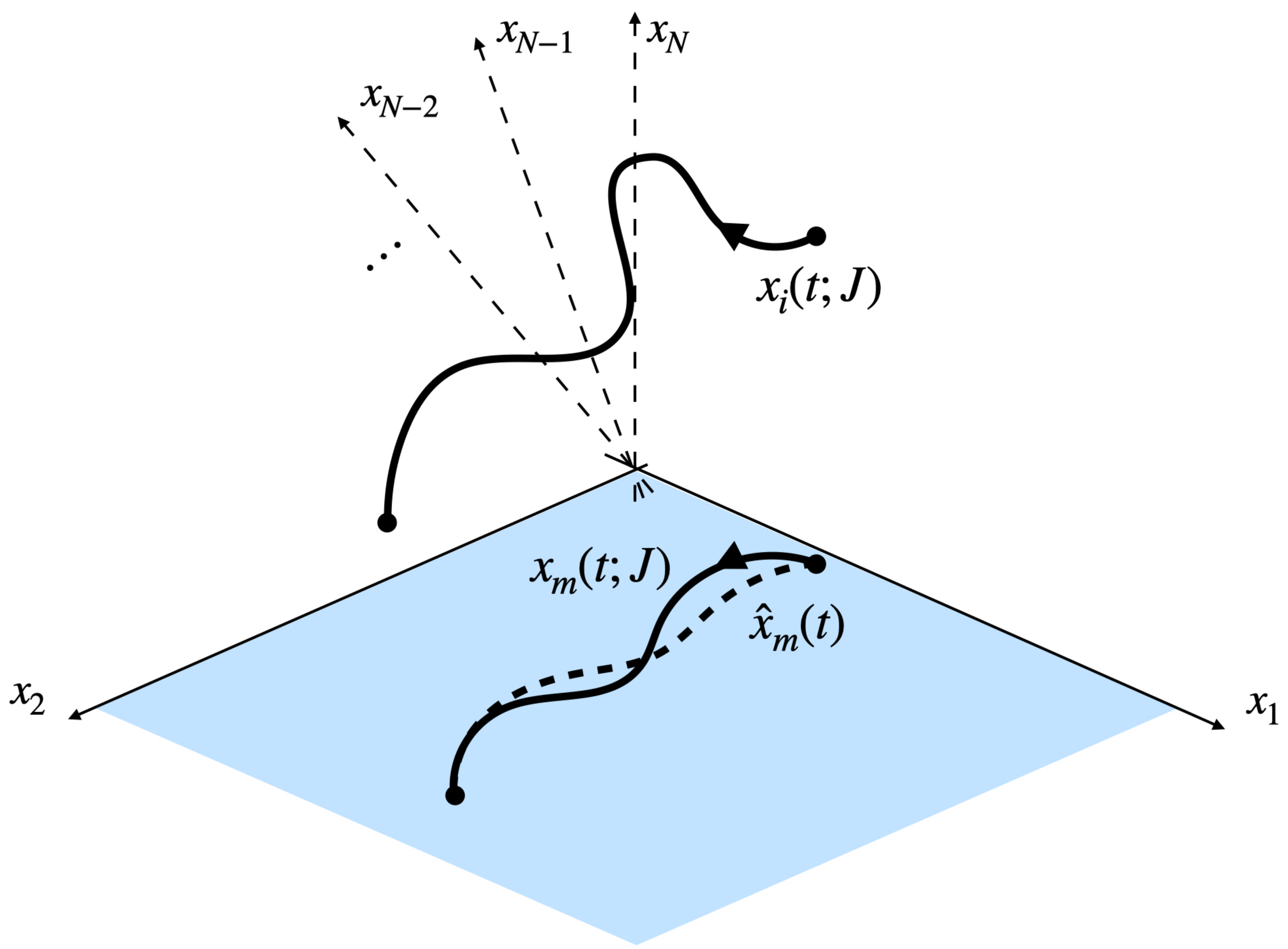}
            \end{minipage}
            \caption{ (A) Network model for developmental dynamics. Each node represents the state of the system and follows Eq.~\eqref{eq: model}. Only a subset of the nodes characterize the developmental trajectory which are colored in the figure. (B) A cartoon of the trajectory is shown here. Based on the inheritable parameters $J$, the trajectory $x_{i}(t; J)$ is generated. The developmental process is characterized by $x_{m}(t; J)$, the projection of the trajectory to an $M$ dimensional subspace, ($M=2$ here for the purposes of visualization). Individuals are evolved to realize a given reference trajectory $\hat{x}_{m}(t)$. }
            \label{fig: model}
      \end{figure}

      Our network model for developmental trajectories is similar to gene regulatory or neural network models used before, where the value of each node represents the state of given internal components that specify a phenotypic state (Fig.~\ref{fig: model}A)~\cite{Mjolsness1991-vr, Salazar-Ciudad2001-qt,Huang2005-py,Kaneko2007-qm,Hopfield1982-df, Buice2013-md}. We construct a network with $N (=40)$ nodes where the dynamics of the state of each node $x_{i}(i = 1, 2, \cdots, N)$ is given by a set of Ordinary Differential Equations (ODEs):

      \begin{align}
            \frac{d x_{i}}{d t}= F\left( \sum_{j=1}^{N}J_{ij}x_{j}\right) - x_{i}, \label{eq: model}
      \end{align}

      Here, $F(z)$ is a sigmoidal function $F(z) = 1 / \{1 + \exp(-\beta z)\}$, where $\beta = 40$. In Eq.~\eqref{eq: model}, the matrix $J$ determines the regulatory relations of the nodes, which can be both activatory or inhibitory. The matrix elements are treated as parameters that are inheritable but can change in evolutionary time (i.e., across generations). The dynamics of the nodes is a function of developmental time and a fixed parameter set $J$, $x_{i}(t; J)$. We introduce a subspace $x_{m}(t; J)$ whose state characterizes the developmental trajectory ($m = 1, 2, \cdots , M = 3$). Hence, in our model, ``development" is the generation of a trajectory based on a particular matrix $J$ and ``evolution" is the slow dynamics of the matrix elements $J$ in a population. We do not introduce any noise in the developmental process.

      Where our model significantly departs from previous such models is in our choice of the fitness function. Rather than optimizing $J$ to reach a fixed end-point, we optimize $J$ to realize a desired developmental trajectory (Fig.~\ref{fig: model}B). A reference trajectory is defined $\hat{x}_{m}(t)$ and the fitness for a given $J$ is defined as the distance, between the generated trajectory $x_{m}(t; J)$ and the reference trajectory $\hat{x}_{m}(t)$. Mathematically,
      \begin{align}
            f(J) = - \frac{1}{T}\int^{T}_{0}\sqrt{\sum_{m=1}^{M}\left( x_{m}(t; J) - \hat{x}_{m}(t) \right)^{2}}d t. \label{eq: fitness}
      \end{align}
      More generally, we can define the distance between two trajectories $x_{m}^{k}( t)$ and $x_{m}^{l}(t)$, $D(x_{m}^{k}, x_{m}^{l})$ as
      \begin{align}
            D\left( x_{m}^{k}(t), x_{m}^{l}(t) \right) = \frac{1}{T}\int^{T}_{0}\sqrt{\sum_{m=1}^{M}\left( x_{m}^{k}(t) - x_{m}^{l}(t) \right)^{2}}d t. \label{eq: distance}
      \end{align}

      To maximize $f(J)$ with a given $\hat{x}_{m}(t)$, we performed an evolutionary simulation. We begin with a population of size $L(=120)$ with $J$ randomly assigned from values $\{-1, 0, 1\}$ with probability $\{1/4, 1/2, 1/4\}$. We calculate the fitness value of each individual based on the generated trajectory. Here, initial conditions for the ODEs are fixed as $x_{m}(t=0) = \hat{x}_{m}(t=0)$, whereas $x_{i}(t =0)$ ($M < i < N$) are assigned random value from a uniform distribution $[0, 1]$. After calculating the fitness, individuals are probabilistically selected to form the next generation population of size $L$. We define the probability of selection for the $l$-th individual with fitness $f_{l}$ as
      \begin{align}
            p(l) = (\exp (- \nu \times f_{l})) / Z,        \\
            Z = \sum_{l'=1}^{L}\exp (- \nu \times f_{l'}),
      \end{align}
      where $\nu=2.0$. Mutations were simulated by replacing elements of the matrix $J$, where a given number $\mu = 2$ elements from the matrix were replaced with new values randomly assigned as before.

      To compare the results of our evolutionary simulation, we also sampled functional individuals by a random sampling method. For efficient sampling, we used the multicanonical Monte Carlo (McMC) method to obtain such individuals~\cite{Nagata2020-el, Wang2001-or, Wang2001-qo} (see SI for more details).

      \subsubsection*{Results}

\begin{figure}[htbp]
            \centering
            \begin{minipage}[c]{0.48\linewidth}
                  \begin{flushleft}
                        A
                  \end{flushleft}
                  \centering
                  \includegraphics[width=\textwidth]{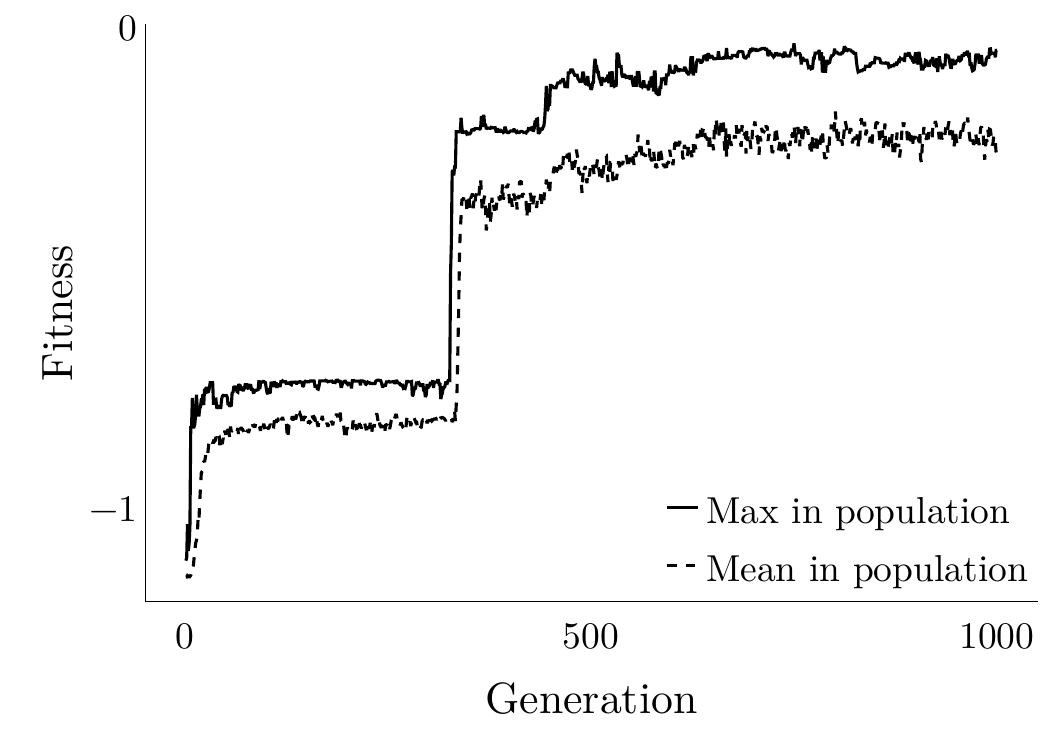}
            \end{minipage}
            \hfill
            \begin{minipage}[c]{0.48\linewidth}
                  \centering
                  \begin{flushleft}
                        B
                  \end{flushleft}
                  \includegraphics[width=\textwidth]{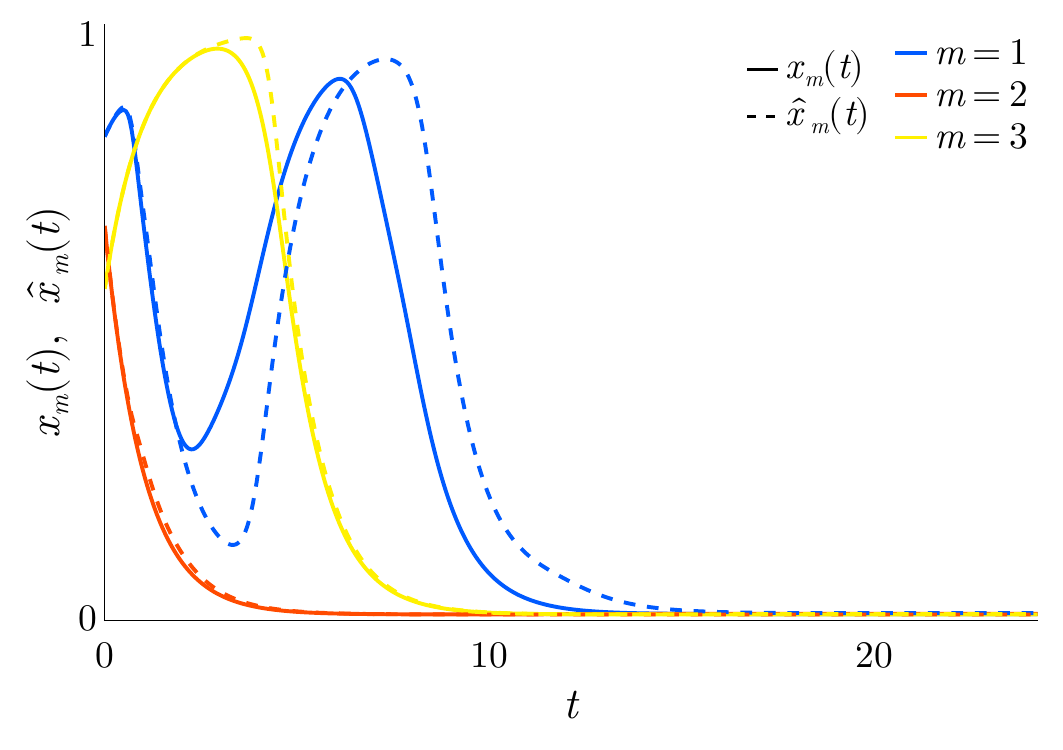}
            \end{minipage}
            \caption{ (A) Example of an evolutionary simulation. We plot maximum and mean fitness value in the population at each generation. (B) Time development of $x_{m}(t; J)$ (solid lines) and $\hat{x}_{m}(t)$ (dotted lines) for an evolved individual. Different nodes are plotted in different colors.}
            \label{fig: evolution}
      \end{figure}

      We conducted our evolutionary simulations and were able to evolve a population whose developmental trajectories were close to the reference trajectory. The value of the fitness as a function of generation is plotted in Fig.~\ref{fig: evolution}A and shows the increase and eventual saturation of fitness. We show a typical trajectory of an evolved individual along with the reference trajectory in Fig.~\ref{fig: evolution}B. We were also able to obtain functional individuals with the desired trajectory from random sampling.

\begin{figure}[htbp]
            \centering
            \includegraphics[width=0.6\textwidth]{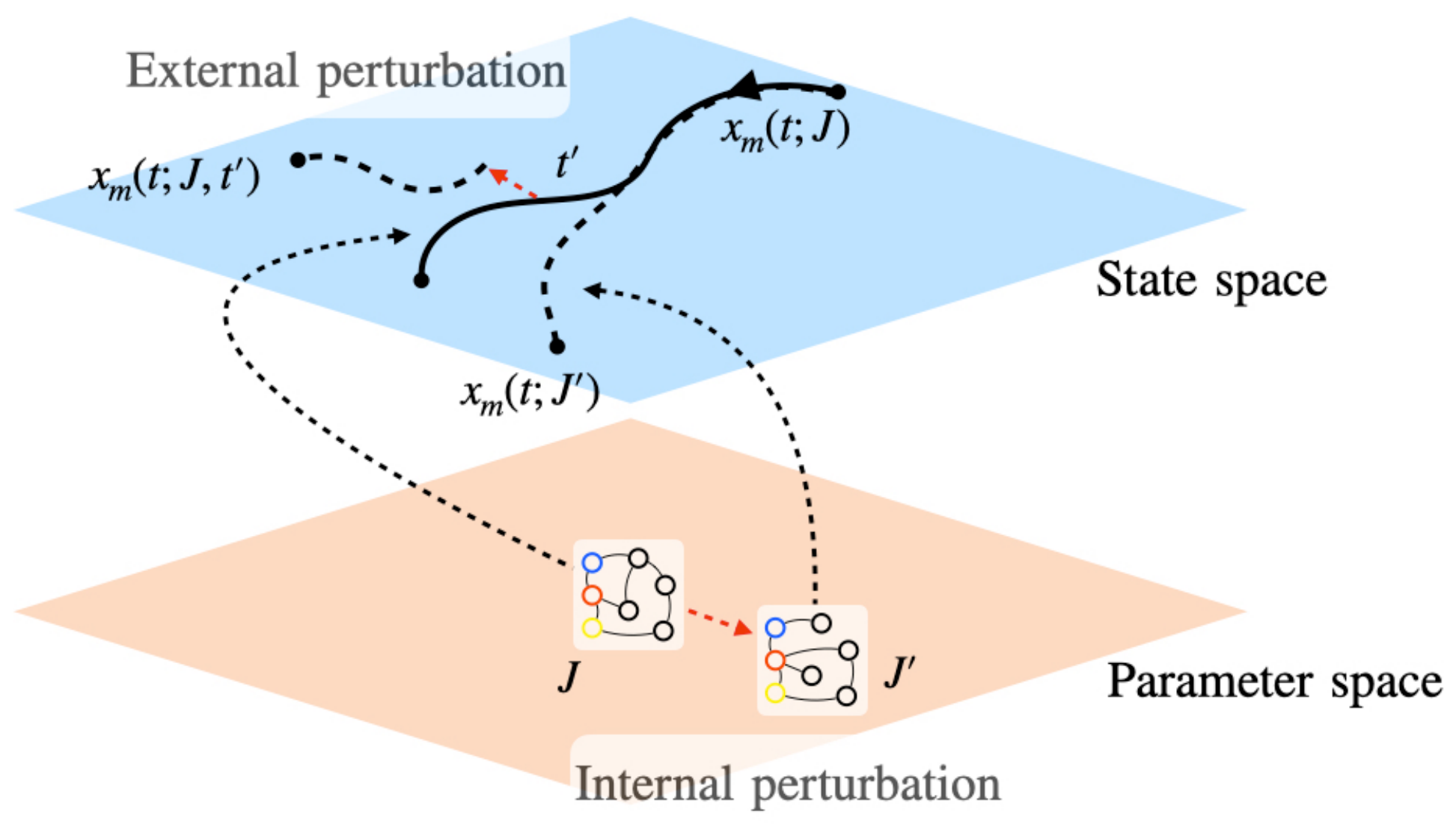}
            \caption{ A cartoon of the two kinds of perturbations considered in the paper. $x_{m}(t; J)$ is the original trajectory with given parameters $J$. External perturbations perturb the phenotypic space (like environmental perturbations), whereas internal perturbations perturb the fixed parameter space (like genetic perturbations or mutations). }
            \label{fig: two perturbations}
      \end{figure}

      To investigate whether the trajectories thus evolved were robust against perturbations, we considered two types of perturbations: external and internal (Fig.~\ref{fig: two perturbations}). External perturbations (e.g. from the environment) randomly perturb the value of $x_{i}$ at a fixed time $t'$ from a uniform distribution $-\Delta$ to $\Delta$ with $\Delta = 0.25$. Internal perturbations (e.g. to the genotype) change the matrix $J$ and hence affect the trajectory. Given a base trajectory $x_{m}(t; J)$, we denote externally perturbed trajectories as $x_{m}(t; J, t')$ where $t'$ is the time of perturbation. Internally perturbed trajectories are denoted by $x_{m}(t; J')$, with mutated genotype $J'$.

\begin{figure}[htbp]
            \centering
            \begin{minipage}[c]{0.48\linewidth}
                  \begin{flushleft}
                        A
                  \end{flushleft}
                  \centering
                  \includegraphics[width=\textwidth]{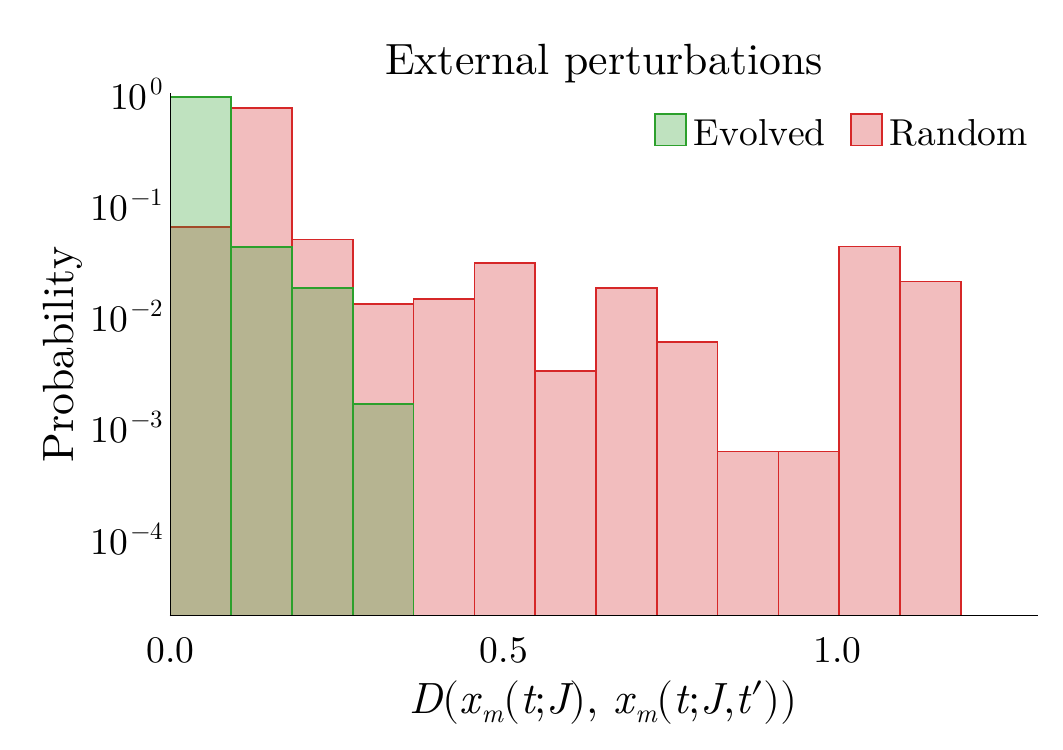}
            \end{minipage}
            \hfill
            \begin{minipage}[c]{0.48\linewidth}
                  \centering
                  \begin{flushleft}
                        B
                        \includegraphics[width=\textwidth]{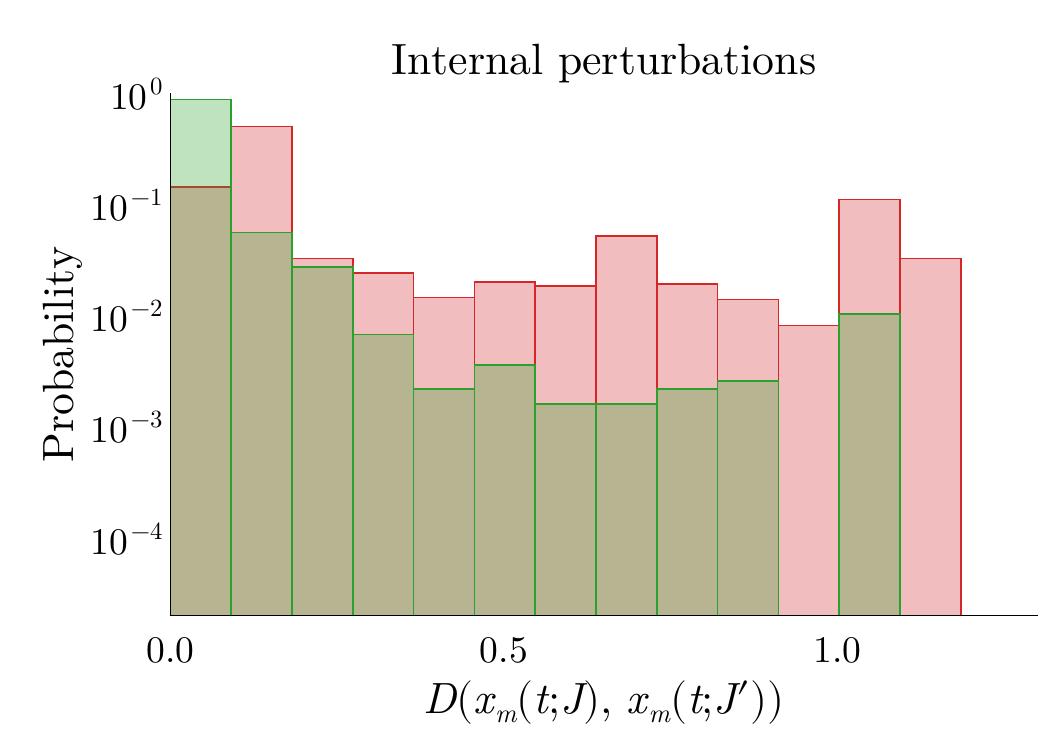}
                  \end{flushleft}
            \end{minipage}
            \caption{ (A) Histogram of the distance between externally perturbed trajectories ${x_m (t; J, t')}$ and the original trajectory $x_{m}(t; J)$. (B) Histogram of the distance between internally perturbed trajectories ${x_m (t; J')}$ and original trajectory $x_{m}(t, J)$. We have overlaid the histogram obtained as a result of these perturbations performed on an evolved individual (green) and a randomly sampled individual (red). }
            \label{fig: distance}
      \end{figure}

      To investigate the effect of evolution on these trajectories, we compared the effect of perturbations on our evolved population with the effects of small perturbation on a randomly sampled population. We generated 5000 externally perturbed trajectories $\{x_{m}(t; J, t')\}$ by perturbing a given trajectory at 50 time points 100 times. We also generated 5000 internally perturbed trajectories $\{x_{m}(t; J')\}$. We measured the distance between the perturbed trajectories and the base trajectory $x_{m}(t; J)$. The histogram for distances for evolved and randomly sampled individuals is plotted in Fig.~\ref{fig: distance}. Interestingly, we obtain greater robustness (as evidenced by reduced distance from the base trajectory upon perturbation) for both external and internal perturbations in the evolved individuals. It should be noted that these individuals were not exposed to any external perturbations during the process of evolution. We hence conclude that our evolved trajectories are canalized to both external and internal perturbations.

      The above results are shown for a given individual but similar effects hold across the population. We see the effects of internal and external perturbations among 200 individuals obtained using evolutionary simulations or random sampling (Fig.~S1) and note that evolution canalizes the effect of both internal and external perturbations. These results indicate that the effect of the two different kinds of perturbations are linked, and evolutionary processes selecting for robustness to one kind of perturbation may also produce robustness in the other kind of perturbation (see SI for more details).

\begin{figure}[htbp]
            \centering
            \begin{minipage}[c]{0.48\linewidth}
                  \begin{flushleft}
                        A
                  \end{flushleft}
                  \centering
                  \includegraphics[width=\textwidth]{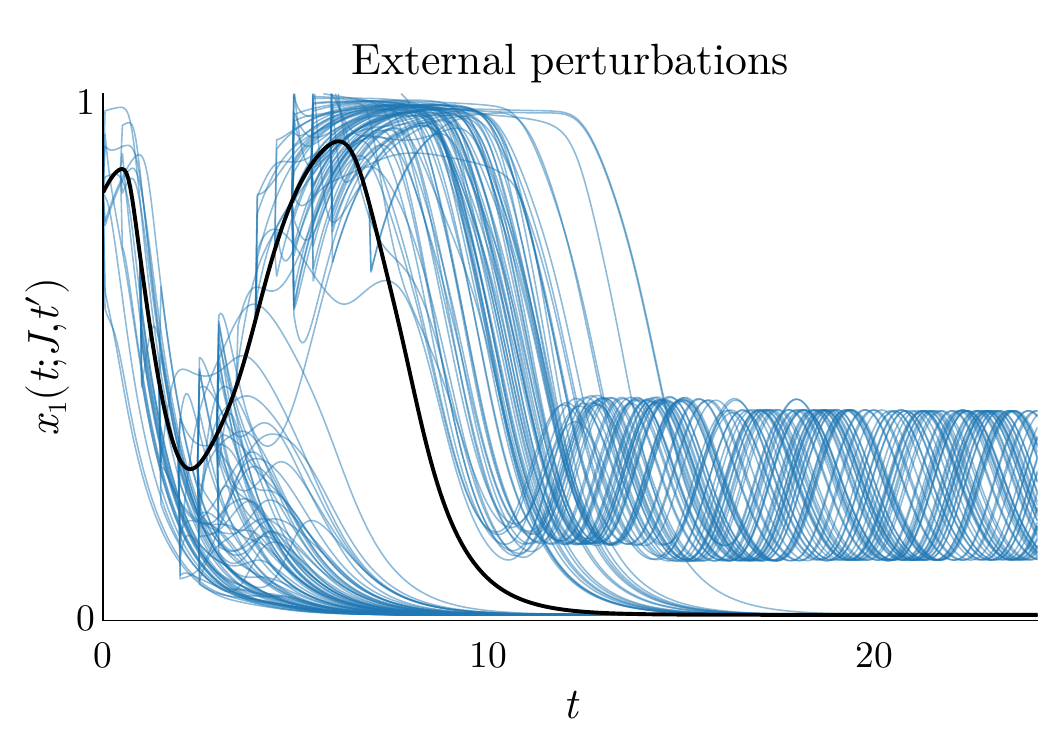}
            \end{minipage}
            \hfill
            \begin{minipage}[c]{0.48\linewidth}
                  \centering
                  \begin{flushleft}
                        B
                  \end{flushleft}
                  \includegraphics[width=\textwidth]{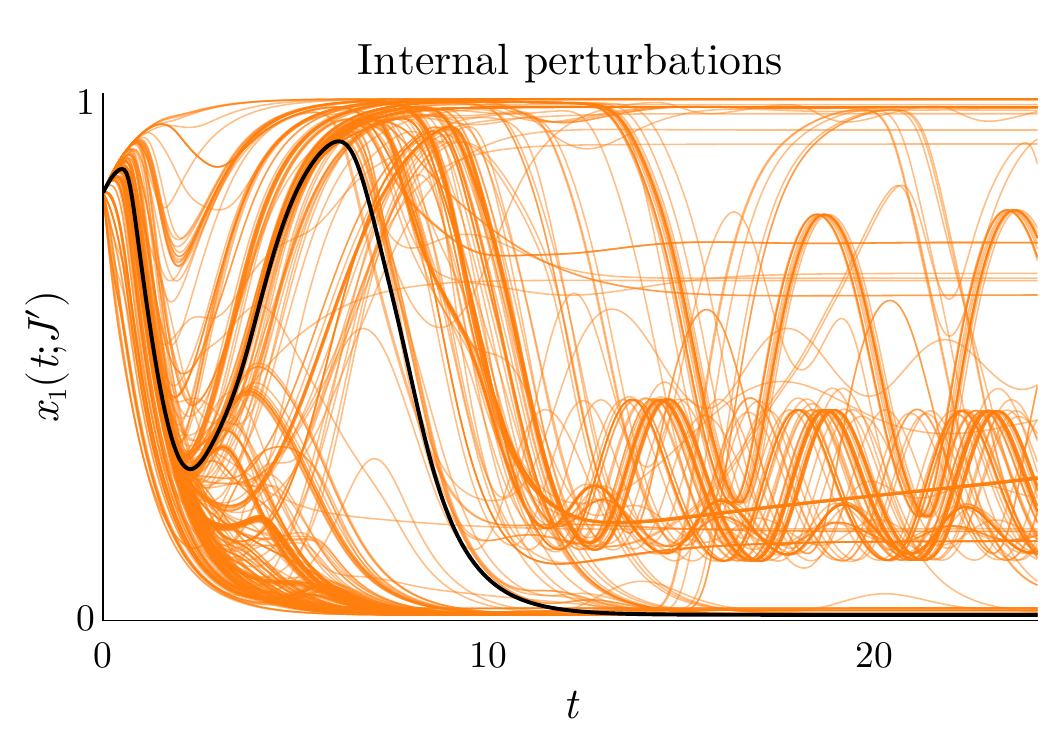}
            \end{minipage}
            \begin{minipage}[c]{0.48\linewidth}
                  \centering
                  \begin{flushleft}
                        C
                  \end{flushleft}
                  \includegraphics[width=0.8\textwidth]{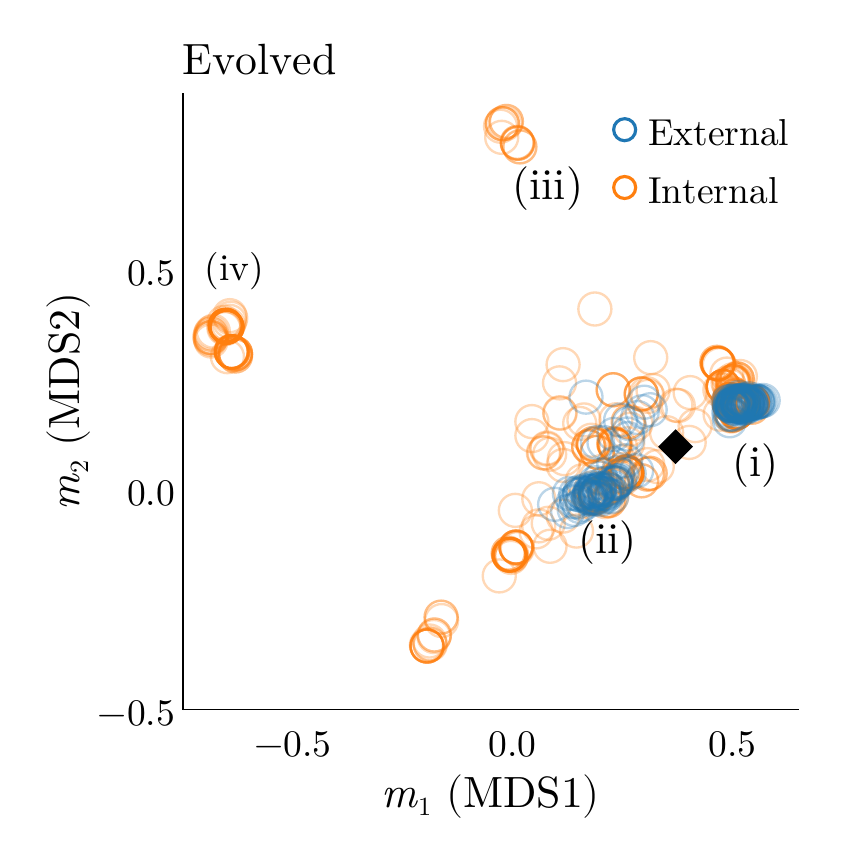}
            \end{minipage}
            \begin{minipage}[c]{0.48\linewidth}
                  \centering
                  \begin{flushleft}
                        D
                  \end{flushleft}
                  \includegraphics[width=0.8\textwidth]{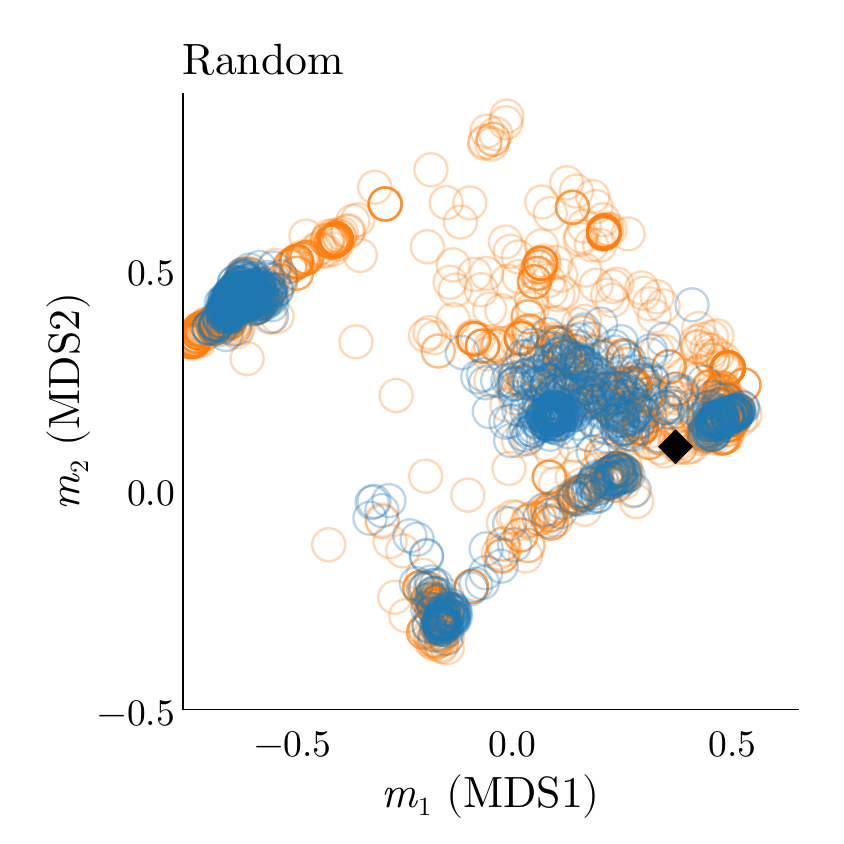}
            \end{minipage}
            \begin{minipage}[c]{0.8\linewidth}
                  \centering
                  \begin{flushleft}
                        E
                  \end{flushleft}
                  \includegraphics[width=\textwidth]{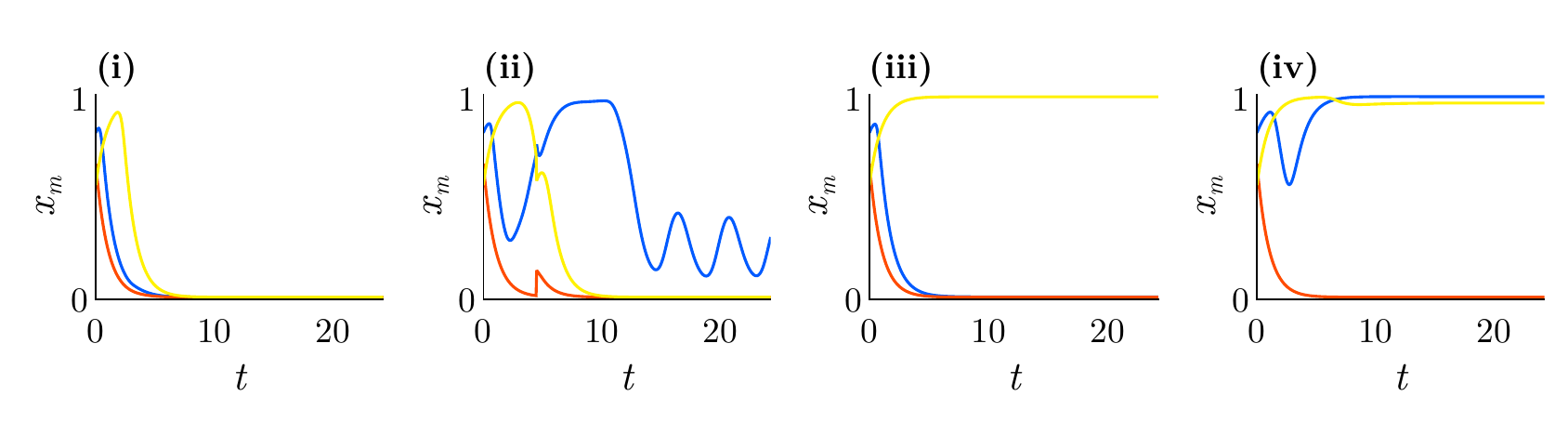}
            \end{minipage}
            \begin{minipage}[c]{0.6\linewidth}
                  \centering
                  \begin{flushleft}
                        F
                  \end{flushleft}

                  \includegraphics[width=\textwidth]{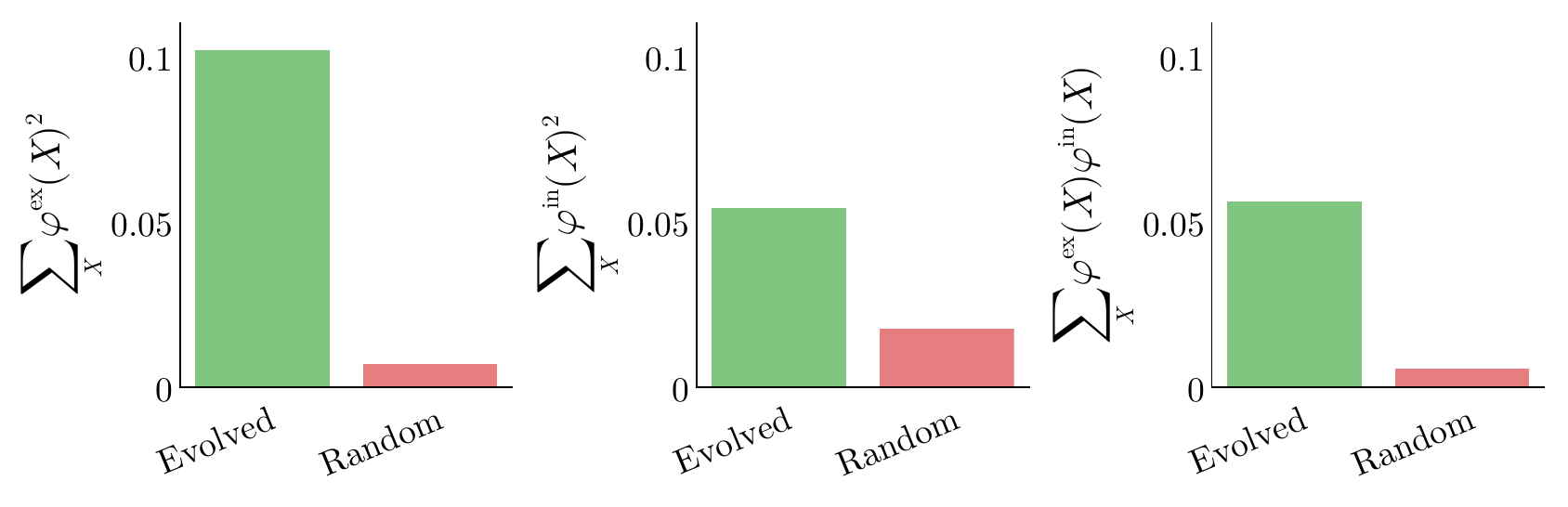}
            \end{minipage}
            \caption{(A-B) The time series of one of the nodes (A) $\{x_{1}(t; J, t')\}$ and (B)$\{ x_{1}(t; J')\}$ on perturbing an evolved individual. (C-D) Two dimensional mapping of $\{x_{m}(t; J )\}$ and $\{x_{m}(t; J')\}$ from (C) an evolved individual and (D) a randomly sampled individual. The black diamond shows the reference trajectory $\hat{x}_{m}(t)$ projected onto the space (see SI for more details). (E) The actual trajectories at each point (i)-(iv) in (C) and (D) demonstrating very distinct but nevertheless overlapping effects of perturbations. (F) Comparison of $\sum_{X}\varphi^{\mathrm{ex}}(X)^{2}$, $\sum_{X}\varphi^{\mathrm{in}}( X)^{2}$, and $\sum_{X}\varphi^{\mathrm{ex}}(X) \varphi^{\mathrm{in}}(X)$, in an individual obtained from evolutionary simulations (green) and random sampling (red).}
            \label{fig: mds}
      \end{figure}

      We next tested the effect of large perturbations. We found alternative developmental trajectories emerge under large perturbations. We only kept those trajectories where the distance from the original trajectory is sufficiently far. Surprisingly, we find the emergence of similar types of alternative trajectories for both internal and external perturbations with large effects as shown in Fig.~\ref{fig: mds}A and Fig.~\ref{fig: mds}B. Though we work with a given reference trajectory, this phenocopying is observed for various different choices of reference trajectories as shown in Fig.~S2.

      To examine this phenomenon in more detail, we utilized MultiDimensional Scaling (MDS) to dimensionally reduce the trajectories~\cite{Borg2005-dp}. The basic principle of MDS is to construct a low dimensional representation which preserves the distance between trajectories as far as possible. It is clear that the lower dimensional representation of trajectories in the evolved individual (Fig.~\ref{fig: mds}C) is much more tightly clustered than in a random individual (Fig.~\ref{fig: mds}D). Further, the overlap between external and internal perturbations is much stronger for a typical evolved individual than for a random individual. We show some representative alternative trajectories in Fig.~\ref{fig: mds}E.

      To quantify these properties, we compared the probability distribution in the MDS space. We kept the top five MDS modes to obtain a distribution $\varphi^{\mathrm{ex}}(X)$ and $\varphi^{\mathrm{in}}(X)$ where the subscripts refer to the externally and internally perturbed trajectories respectively. Here, $X =\{X_{1}, X_{2}, ..., X_{5}\}$, where $X_{k}$ is the $k$-th MDS mode. As a measurement of the amount of localization of the distribution, we calculated the sum of the square of the distribution $\sum_{X}{\varphi^{\mathrm{ex}} (X)^2}$ and $\sum_{X}{\varphi^{\mathrm{in}} (X)^2}$. This quantity takes higher values when the distribution is localized and lower values when the distribution is uniform. To measure the extent of overlap of the two distributions, we calculated $\sum_{X}{\varphi^{\mathrm{ex}} (X) \varphi^{\mathrm{in}} (X)}$. This quantity takes higher values when the two distributions are close together and lower values when they are not. As we show in Fig.~\ref{fig: mds}F, both of these quantities are significantly higher for evolved trajectories. The localization of the distributions is reflective of the canalization of these trajectories. The overlap between externally and internally perturbed trajectories is an indication that the property of phenocopying has evolved even though it is not selected for (see Fig. S3 for a statistical comparison).

\begin{figure}[htbp]
            \centering
            \begin{minipage}[c]{0.48\linewidth}
                  \begin{flushleft}
                        A
                  \end{flushleft}
                  \centering
                  \includegraphics[width=0.8\textwidth]{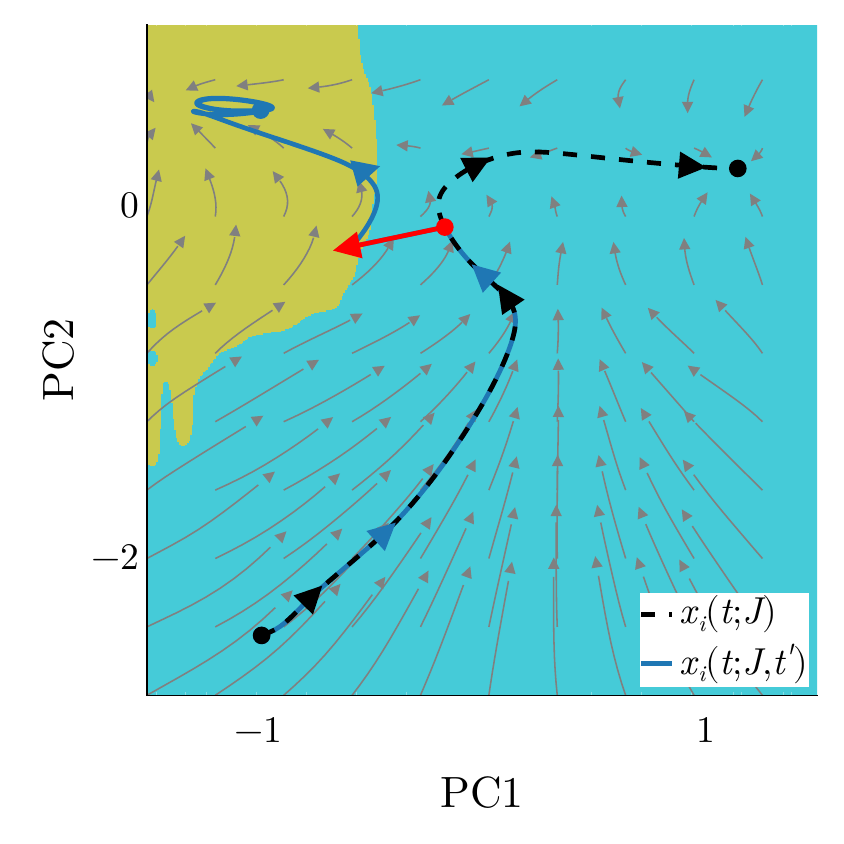}
            \end{minipage}
            \hfill
            \begin{minipage}[c]{0.48\linewidth}
                  \begin{flushleft}
                        B
                  \end{flushleft}
                  \centering
                  \includegraphics[width=0.8\textwidth]{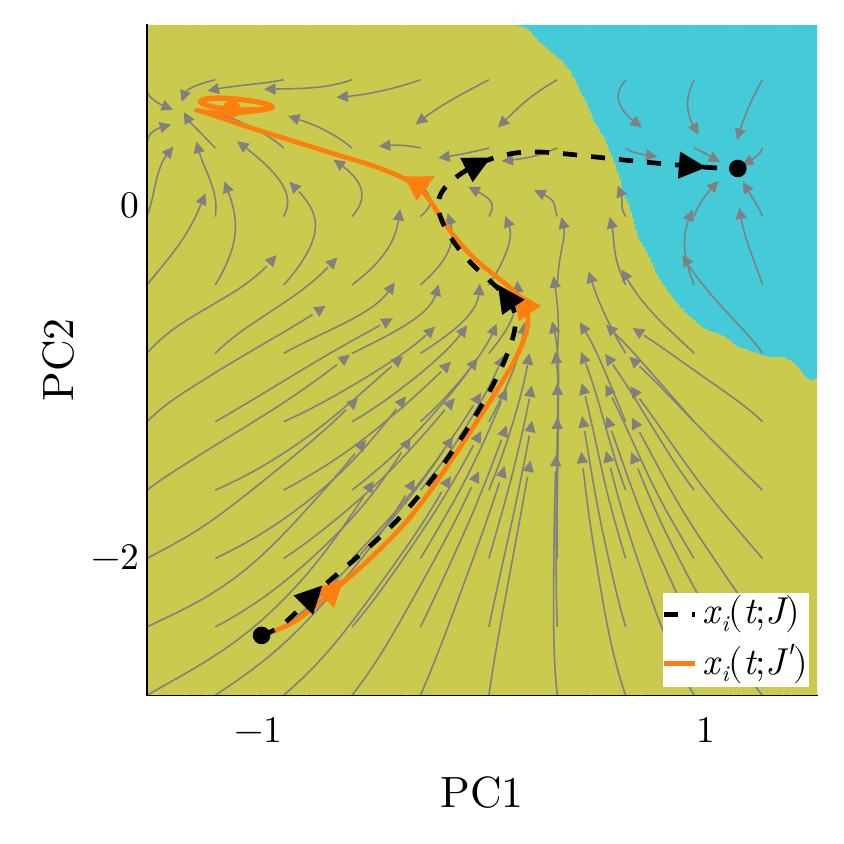}
            \end{minipage}
            \begin{minipage}[c]{0.48\linewidth}
                  \begin{flushleft}
                        C
                  \end{flushleft}
                  \centering
                  \includegraphics[width=0.8\textwidth]{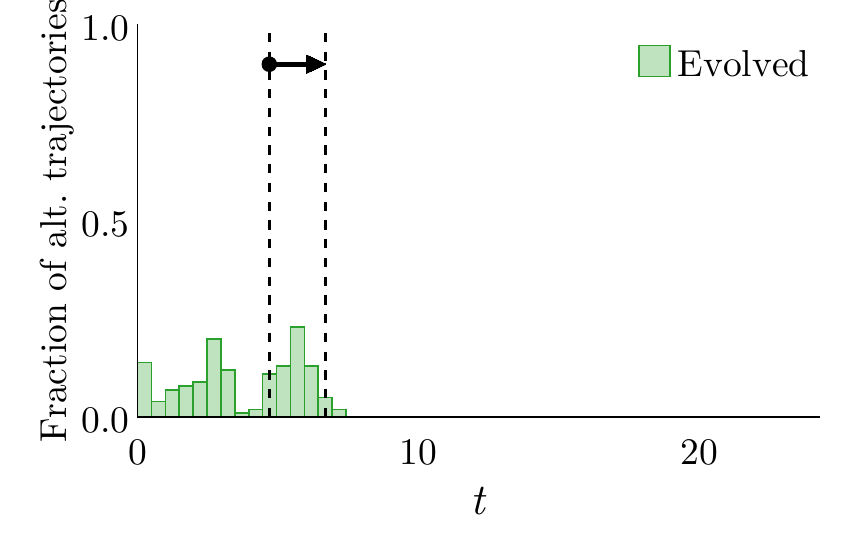}
            \end{minipage}
            \hfill
            \begin{minipage}[c]{0.48\linewidth}
                  \begin{flushleft}
                        D
                  \end{flushleft}
                  \centering
                  \includegraphics[width=0.8\textwidth]{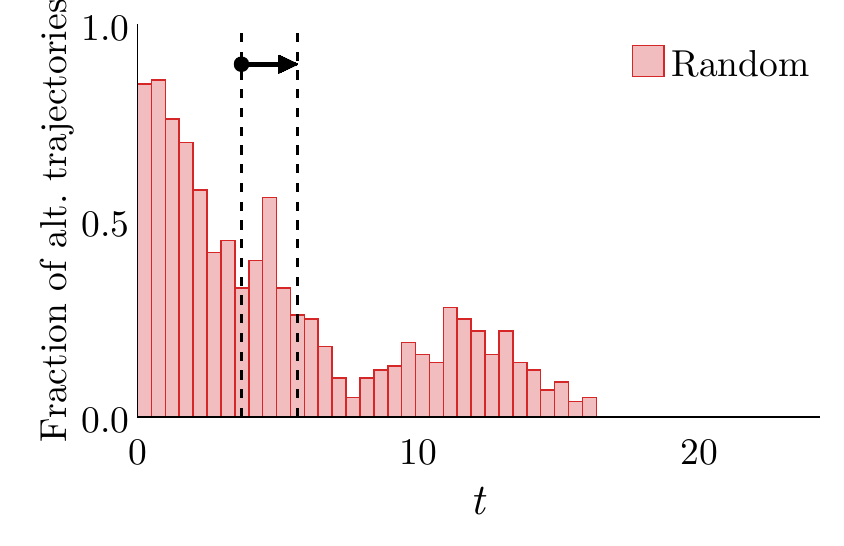}
            \end{minipage}
            \begin{minipage}[c]{0.48\linewidth}
                  \begin{flushleft}
                        E
                  \end{flushleft}
                  \centering
                  \includegraphics[width=0.8\textwidth]{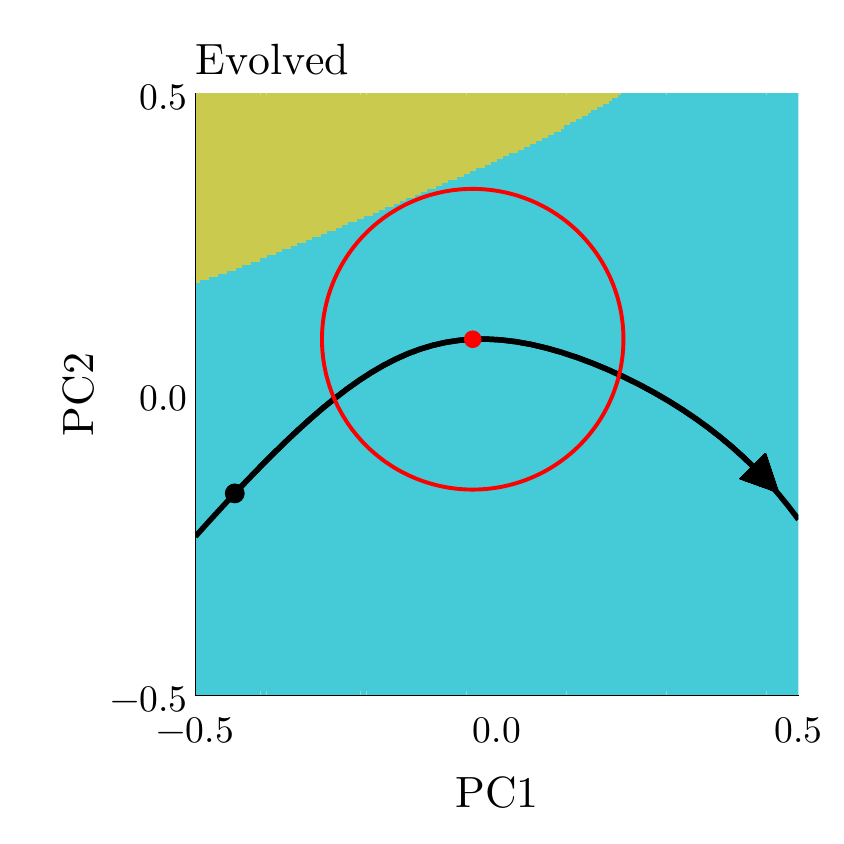}
            \end{minipage}
            \hfill
            \begin{minipage}[c]{0.48\linewidth}
                  \centering
                  \begin{flushleft}
                        F
                  \end{flushleft}
                  \includegraphics[width=0.8\textwidth]{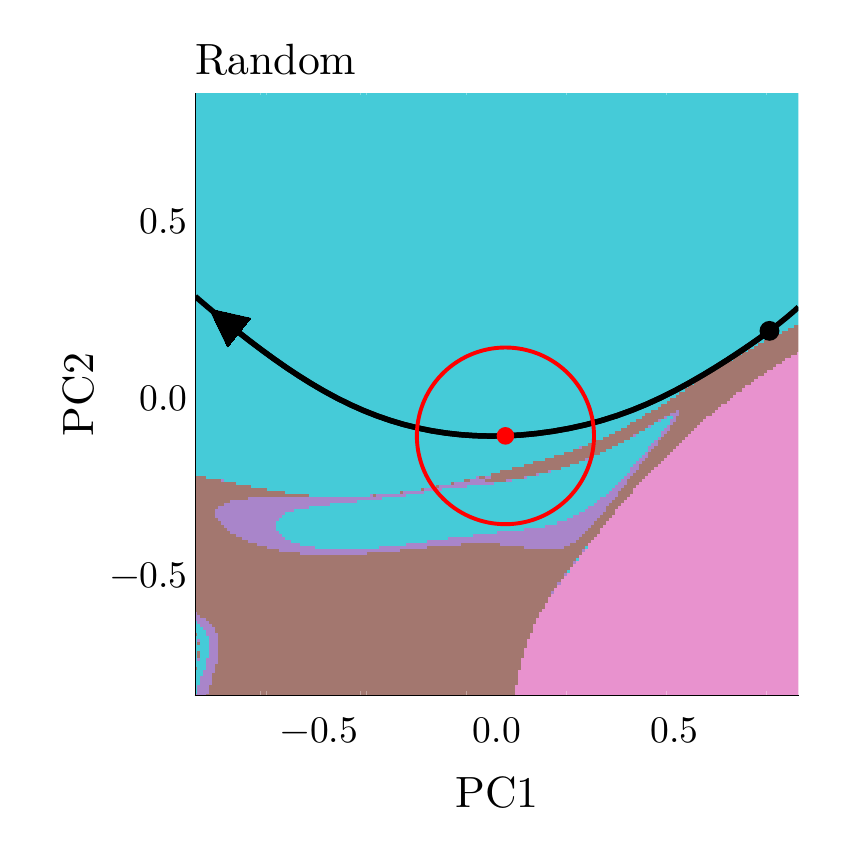}
            \end{minipage}
            \caption{ (A-B) Trajectories projected on their PC axis. The original trajectory is shown as black and dashed. (A) The external perturbation shifts the trajectory $x_{i}(t; J, t')$ as indicated by the red arrow leading to a distinct trajectory shown in blue. (B) The trajectory obtained as a result of changing the parameters $J$, $x_{i}(t; J')$ shifts the basin structure leading to a very similar trajectory as in (A) shown in red. (C-D) Sensitivity against environmental perturbations as a function of time $t$, in an evolved individual and randomly sampled individual. We randomly and externally perturb the trajectory at each timepoint 100 times, and calculate the fraction of perturbations leading to an alternative trajectory. (E-F) Comparison of the phase portrait of an evolved individual and a random sampled individual. The red point represents the timepoint with the peak in Fig.~\ref{fig: phaseportrait}C and D, and the red circle represents the region in which external perturbations push out.}
            \label{fig: phaseportrait}
      \end{figure}

      To explain these observations, we tried to understand them from a dynamical systems viewpoint. In general, the basin structure is complex and high dimensional. For simplicity, we focused on the structure of the flow near given trajectories and projected the flow on to the principal components (PCs)~\cite{Pearson1901-ms}. In Fig.~\ref{fig: phaseportrait}A, we plot the flows of a single individual. The basins are also projected on to this space. Note that this is a lower dimensional projection of a higher dimensional flow. An external perturbation at a given time can shift the trajectory and take it to an alternative basin. However, the strength of the external perturbation needs to be sufficient to cross the boundaries between the two basins. The trajectory remains robust to small perturbations. In Fig.~\ref{fig: phaseportrait}B, we see the effect of an internal perturbation which shifts the basin structure but does not strongly affect the flow.

      To compare the phase portrait between an evolved and a randomly sampled individual, we quantified the sensitivity to external perturbations as a function of developmental time. We did this by calculating the fraction of perturbations which lead to an alternative trajectory out of 100 perturbations at each timepoint. In Fig.~\ref{fig: phaseportrait}C and Fig.~\ref{fig: phaseportrait}D, we see that this fraction is strongly suppressed in the evolved case as compared to the random case at nearly all time-points. If we focus on the most sensitive time-points and look at the basin structure in the PC space at nearby times, we find that the basin structure of the evolved individual is considerably simplified than the randomly sampled individual (Fig.~\ref{fig: phaseportrait}E and F). It is therefore more plausible that external and internal perturbations lead to similar effects in the evolved case, whereas the complex basin structure in the random case implies that external and internal perturbations will typically have distinct effects. Such differences in the basin structure difference could be observed not only locally near a sensitive point, but also globally along the whole trajectory (Fig.~S4).

      \subsubsection*{Discussion}

      We have numerically demonstrated that the phenomenon of phenocopying can emerge with an abstract evolutionary model of developmental systems. To investigate this phenomenon, we evolve entire developmental trajectories rather than a fixed end-point phenotype. Therefore, our fitness is a function of the entire trajectory rather than simply the final phenotype. Such dynamic fitness is usually accounted for using age or stage dependence.

      By explicitly accounting for perturbations that could be internal (e.g. genetic) or external (environmental), we show that evolving for robustness to internal perturbations can automatically create robustness to external perturbations. Our simulations therefore are in accordance with the hypothesized single mode of canalization suggesting that the strongly canalized parts of developmental trajectories will be robust to both environmental and genetic perturbations~\cite{meiklejohn2002single}. However, we note that this ``mode" may not be a single molecule and could be context dependent~\cite{takahashi2019multiple}. Our work contributes to the literature on how the effects of environmental and genetic perturbations may be related by investigating it in an explicitly dynamic scenario.

      The phenomenon of phenocopies has a large experimental literature from the early days of embryology. However, it has received little theoretical attention. In our framework, we define phenocopies as alternative developmental trajectories which can be produced both by large external perturbations to the state of a system or by changing internal parameters. We find that these alternative trajectories emerge from our evolutionary simulations without being selected for. Our simulations evolve the system to follow a fixed reference trajectory. Importantly, randomly sampled populations which have not undergone evolution but whose dynamics is close to the reference trajectory do not have the property of phenocopying. This general idea of phenocopies as the identical effects resulting from large external and internal perturbations may have application in other very different contexts~\cite{Furusawa2018-fi, feng2023activity, Russo2025-ks, Chuang2019-by}

      More modern experimental work is needed to understand phenocopies. A recent study suggested that general stress-response mechanisms may be responsible for phenocopies in the case of the bithorax phenocopy induced by ether or heat stress~\cite{snir2024organ, elgart2015stress}. In general, however, little data exists on comparing gene expression pattern changes under environmental perturbations in developmental systems. Another study observed that applying directional selection on wing size for flies decanalizes the system simultaneously to both genetic and environmental perturbations, which is consistent with our results that show a correspondence between the effect of the two kinds of perturbations~\cite{groth2018directional}.

      Waddington's experiments on genetic assimilation, which have recently been repeated and studied~\cite{fanti2017canalization, marzec2022reexamining, Sabaris2025-bw}, relied on the presence of phenocopies. His landscape was essential in providing an explanation for genetic assimilation. A recent study theoretically studied genetic assimilation but assumed perfect canalization of both the original and perturbed trajectory, which is not true in general~\cite{Raju2023-ik}. Therefore, our work may provide further insight into the phenomenon of genetic assimilation which has found application across species \cite{van-der-Burg2020-ex, vigne2021single}.

      In the study of robustness and plasticity, the dynamical viewpoint of development can be very useful. However, since developmental trajectories are in high dimensional space, the structure of their phase space can be quite complex~\cite{zhang2021basins}. Some have suggested that linearizing the dynamics and looking at the structure of the modes may provide insight into why genetic and environmental perturbations have similar effects because they are mediated through the "soft-modes"~\cite{Russo2025-ks}. However, the phenomenon of phenocopies involves large perturbations and is inherently non-linear. Though it is more difficult to provide general results in this case, our numerical results suggest that this phenomenon emerges because evolution simplifies the complex basin structure of high dimensional systems. This is reminiscent of the observed low-dimensional structure of high-dimensional systems observed in other contexts and might be regulated by "slow variables" as in~\cite{Kaneko2024-ag}. Further theoretical studies are required to understand this.

      \section*{Acknowledgements}
      We acknowledge support from the Department of Atomic Energy, Government of India (under project RTI4006) and the Simons Foundation (287975). YM thanks Kunihiko Kaneko for insightful discussions. AR acknowledges helpful discussions with BingKan Xue and Stanislas Leibler. We thank Kunihiko Kaneko and Hiroshi Hamada for reading a draft of this manuscript and giving helpful suggestions.

      \bibliographystyle{unsrt}
      \bibliography{reference}
\end{document}